\documentclass[a4paper]{article}
\usepackage{lmodern}

\usepackage[T1]{fontenc}
\usepackage[latin9]{inputenc}
\usepackage{amsthm}
\usepackage{amsmath}

\makeatletter

\newcommand{\noun}[1]{\textsc{#1}}

\newenvironment{lyxcode}
{\par\begin{list}{}{
\setlength{\rightmargin}{\leftmargin}
\setlength{\listparindent}{0pt}
\raggedright
\setlength{\itemsep}{0pt}
\setlength{\parsep}{0pt}
\normalfont\ttfamily}%
 \item[]}
{\end{list}}
\newcommand{\code}[1]{\texttt{#1}}

\makeatother

\usepackage{listings}

\begin{document}
\newcommand{\NOOP}{\mathbf{NOOP}}
\newcommand{\GNOOP}{\mathbf{GNOOP}}

\title{Towards an Accurate Mathematical Model\\
of Generic Nominally-Typed OOP}

\author{\smallskip{}
	Moez A. AbdelGawad\\
	College of Mathematics, Hunan University\\
	\smallskip{}
	Changsha 410082, Hunan, P.R. China\\
	\smallskip{}
	Informatics Research Institute, SRTA-City\\
	\smallskip{}
	New Borg ElArab, Alexandria, Egypt\\
	\texttt{moez@cs.rice.edu}}

%
%

\maketitle
\abstract{The construction of $\GNOOP$ as a domain-theoretic model of generic
	nominally-typed OOP is currently underway. This extended abstract presents
	the concepts of `nominal intervals' and `full generification' that
	are likely to help in building $\GNOOP$ as an accurate mathematical model of
	generic nominally-typed OOP.  The abstract also presents few related 
	category-theoretic suggestions. The presented concepts and suggestions are
	particularly geared towards enabling $\GNOOP$ to offer a precise and simple 
	view of
	so-far-hard-to-analyze features of generic OOP such as variance annotations
	(e.g., Java wildcard types) and erased generics (e.g., Java type erasure).}
\section{Extended Abstract}
Nominally-typed OO languages are among the top most-used programming
languages today. Examples of nominally-typed OO languages include
industrial-strength mainstream OO programming languages such as Java~\cite{JLS14},
\noun{C\#~\cite{CSharp2015},} \noun{C++}~\cite{CPP2011}, and Scala~\cite{Odersky14}.
Recently, we presented $\NOOP$ as a domain-theoretic model of (non-generic) 
nominally-typed
OOP~\cite{NOOP,NOOPbook,NOOPsumm} and compared it to well-known
structural models of OOP~\cite{AbdelGawad2015a,AbdelGawad2016},
proving---contrary to the mantra ``inheritance is not subtyping''---that
type inheritance and OO subtyping are in full one-to-one correspondence
in nominally-typed OOP~\cite{InhSubtyNWPT13}. To support the development
of $\NOOP$, we also illustrated the technical and semantic value
of nominal typing and nominal subtyping to mainstream OO developers
and language designers~\cite{AbdelGawad2015}.

Generic types add to the expressiveness of type systems of nominally-typed
OO programming languages~\cite{Bank96,Bracha98,Corky98,JLS05,CSharp2007,bloch08,
	ScalaWebsite,JLS14,GenericsFAQWebsite,Zhang:2015:LFO:2737924.2738008}.
Generics provide OO class designers the ability to abstract their
classes over some types, and thus define them ``generically,'' independent
of particular instantiations that class clients may later use. Generics
move the decision as to what actual types to be used for some of the
types used inside the class to the usage-sites of a class (\emph{i.e.},
are decided by the clients of the class) rather than be declared and
fixed at declaration-sites (\emph{i.e.}, decided by class designers).

Generics also offer OO programmers more flexibility, given that \emph{different}
type parameters of a generic class can be used at different usage
sites, even in the same program. Without generics, such a capability
could only be simulated by a cooperation between class designers and
their clients, depending on OO subtyping and using the so-called ``generic
idiom.'' Using the generic idiom is not a type-safe alternative to
generics, given that using it involves requiring clients to insert
downcasts by hand. Because they circumvent the type system, programs
with downcasts can be type unsafe~\cite{FJ/FGJ}.

We believe building a domain-theoretic model of \emph{generic} nominally-typed
OOP along the lines of $\NOOP$ (\emph{e.g.}, as in~\cite{AbdelGawad2016a})
will offer better chances for having a deeper and more solid understanding
of features of generic mainstream OO languages, such as Java erasure,
variance annotations (including Java wildcards), polymorphic methods,
generic type inference and so on. It will also demonstrate the utility
and importance of including nominal type information in mathematical
models of OOP.

The plenty of research done on OO generics ~(\cite{Bank96,Bracha98,Corky98,JLS05,JLS14,CSharp2007,bloch08,ScalaWebsite,
	GenericsFAQWebsite,Zhang:2015:LFO:2737924.2738008})
has proven that generics is a complex feature to analyze. Features
such as Java wildcards~\cite{Torgersen2004}, in particular, while
designed to ameliorate the conceptual mismatch between parametric
polymorphism and OO subtyping polymorphism, have proven to be difficult
to accurately model~\cite{MadsTorgersen2005,Cameron2007,Cameron2008,Summers2010}.

Based on our ongoing effort to build a mathematical model of OO generics,
we believe a crucial reason for PL researchers not accurately modeling
generics is their use of structural models of OOP. We demonstrate
this here by summarizing our explorations towards constructing a simple
yet precise mathematical model of OO generics that is based on a nominal
rather than structural view of OOP.

Along the lines of our construction of $\NOOP$, we are currently constructing
$\GNOOP$ to be a domain-theoretic\footnote{An introduction to domain theory 
	is presented in
\cite{DomTheoryIntro}.} model of OOP that includes full nominal type
 information found in generic nominally-typed OOP.

The main nominal construct in $\NOOP$ are \emph{class signatures} (see ~\cite{NOOP,NOOPbook,NOOPsumm}). Class signatures in $\NOOP$ include full 
type name information (\emph{a.k.a.}, nominal information) found in 
non-generic nominally-typed OOP. For $\GNOOP$, the main nominal construct 
are \emph{class signature constructors}. When applied to type arguments,
 class signature constructors get instantiated
to (ground) class signatures. (Appendix~\ref{sec:Equations-for-Generic}
presents the formal definition of generic class signatures.) Based on the
 definition of ground signature names and ground signatures, the construction
of $\GNOOP$ goes along similar lines of constructing $\NOOP$.
 (See~\cite{AbdelGawad2016a}
for more details on generic class signatures and on the construction
of $\GNOOP$.)

As constructed, $\GNOOP$ however does not (yet) model some important
features of generic OOP, such as polymorphic methods, bounded type
variables and variance annotations (\emph{e.g.}, Java wildcards).

We discuss how to model polymorphic methods in~\cite{AbdelGawad2016a}.
For the accurate modeling of bounded type variables and wildcards/variance
annotations, we are considering adopting a \emph{unified} approach
to both features, using the notions of `nominal (or, named) intervals'
and `full generification'. We discuss both notions below.

Variance annotations in generic OOP are designed to combine the benefits
and abstraction power of parametric polymorphism (\emph{i.e.}, that provided
by plain generics, with no variance) with the benefits and abstraction power
of OO subtyping polymorphism. Variance annotations allow different
instantiations of generic types to be included in the subtyping relation.
(See~\cite{Igarashi02onvariance-based} for a more detailed discussion
of variance annotations.) Variance annotations come in two main flavors
in mainstream generic OOP: \emph{usage-site} variance annotations
(used \emph{e.g.,} in Scala and C\#) and \emph{declaration-site} variance
annotations (used \emph{e.g.}, in Java; more well-known as Java wildcards).

A critical observation to be made early on when analyzing variance
annotations is that \emph{infinite} chains of supertypes could now
occur in the subtyping relation. For example, given the declaration
of class \code{Enum} in Java, and with a typical class declaration
such as \code{class C extends Enum<C>}, consider the chain of Java
types \code{C}, which has \code{Enum<C>} as a supertype (given the
declaration of class \code{C}), which, by subtyping rules for wildcard
types, has supertype \code{Enum<?~extends C>}, which in turn using
the same rules has \code{Enum<?~extends Enum<C>\textcompwordmark{}>}
as a supertype, which has \code{Enum<?~extends Enum<?~extends C>\textcompwordmark{}>\textcompwordmark{}>}
as a supertype, which has \code{Enum<?~extends Enum<?~extends Enum<C>\textcompwordmark{}>\textcompwordmark{}>\textcompwordmark{}>}
as a supertype, ... and so on. (Did you note the repeating pattern?)

In fact, building on this observation we further made the observation
that (the graph of) the subtyping relation in Java (or in generic nominally-typed
OOP, more generally)---when few artificial restrictions on wildcards are
 relaxed, \emph{e.g.},
them having both lower and upper bounds---is \emph{a fractal}. Fractals,
as graphs, are characterized by having `self-similarity'.\label{lbl:selfsim}
 Due to
invariant subtyping, covariant subtyping and contravariant subtyping
rules, we noted that \emph{three }``transformed'' copies of the
subtyping relation are embedded inside the relation (See~\cite{AbdelGawad2014b}
for more details and illustrations.) While this fractal observation
may not be surprising, given the inductive definition of subtyping,
the intricacy in generic OO subtyping comes when noting the three
transformations of the subtyping relation resulting from having three
forms of subtyping rules between generic types.

As mentioned earlier, to model bounded type variables and variance
annotations we consider using what we call `nominal intervals' and
`full generification', where
\begin{itemize}
\item a \emph{nominal interval} is a type variable name with both upper
and lower bounds where the lower bound is guaranteed to be a subtype
of the upper bound (nominal intervals will model bounded type parameters),
and
\item \emph{full generification} means that \emph{all} types inside a class 
(\emph{e.g.}, as the type of a field, the type of a method parameter/return value, 
or passed as a type argument inside a parameterized type---including wildcard types
and nested type arguments) get replaced by (\emph{a.k.a.}, are ``captured'' into)
new additional \emph{synthetic} type parameters of the class. (Appendix~\ref{sec:Fully-Generified-Generic-Class}
presents the formal definition of fully-generified generic class signatures.
Appendix~\ref{sec:Full-Generification-Code-Examples} presents illustrating code
examples).
\end{itemize}
Existential types, originally from mathematical logic, are used to model wildcards
in virtually all research on generic (\emph{a.k.a.}, polymorphic) OOP.
We believe nominal intervals are simpler than existential types as models of
wildcards.  The notion of intervals we use is a simple generalization of the
notion of intervals over total orders (\emph{e.g.}, over real numbers or integers)
to intervals over partial orders (namely, over the subtyping relation).  Naming
(\emph{i.e.}, giving names to) intervals
regains (and also nicely conceals) the existential nature of wildcards as intervals.

As to full generification, because of their capturing in synthetic type variables, a type argument
inside a fully generified class will \emph{always} be a type variable
 (\emph{i.e.},
the name of a nominal interval, like any other type inside the class)
not a paremeterized type. As such, full generification results in what
can be called `single-nested generics', where 
in the code of a fully-generified class there will be no explicit
multi-level nesting of type arguments (\emph{e.g.}, as in the type
\code{List<List<String>\textcompwordmark{}>}, or in the supertype
\code{Equatable<List<Equatable<E>\textcompwordmark{}>\textcompwordmark{}>}
where \code{E} is a type parameter of class \code{Equatable}~\cite{Greenman2014}).
Multi-level nesting of type arguments in fully-generified code will
only be implicit and indirect, expressed always via (original or synthetic) 
type variables.

In addition to our approach being simple and intuitive, based on preliminary analysis 
we made we expect this approach---based on nominal
intervals and full generification---to particularly allow better, more
accurate modeling of Java \emph{wildcards} and Java \emph{erasure}---two features
of generics that, as we hinted above, have not been modeled accurately
so far.

For wildcards, it should be noted that full generification results in the
``disappearance'' of wildcard types.  Due to capturing them as (synthetic)
type variables/parameters---which are modeled as nominal intervals---full
generification results in a uniform treatment of wildcard types and type
variables.

For type erasure, full generification makes \emph{all} (non-type-variable)
types inside a class to be \emph{explicitly} expressed in the type parameters
clause of the class.  Doing a syntactic comparison on the class code, namely
between its legacy non-generic version and its generic version, a \emph{raw type} (the erased version of a generic type) can then be modeled by some instantiation of the
fully generified generic class.  The code comparison will reveal the type
arguments to be used for the \emph{synthetic} type parameters, which were
not explicitly included in the original generic version of the class.

Finally, to possibly further simplify and strengthen our model, we
are considering using some category theoretic tools in our construction
of a mathematical model of generic nominally-typed OOP. While category
theory has been recently getting applied, at an increasing rate, in
other scientific disciplines~\cite{lawvere2009conceptual,awodey2010category,
	spivak2014category},
the use of category theory in computer science, and in defining the
semantics of programming languages in particular, is well known. It
is well known, for example, that \emph{cartesian-closed} categories
(CCCs) provide accurate models of typed lambda calculus~\cite{CatRecDomEqs82,
	DomTheoryIntro}.
Further, a fruitful point of view takes the objects of a category
to be the types of a programming language~\cite{Pierce91}. In such
a view, type constructors (\emph{i.e.}, generic classes, in an OOP
language) are seen as \emph{functors} between categories. F-bounded
generics (where type variables of generic classes are used in their
own bounds~\cite{Baldan1999}) can then be interpreted using \emph{F-coalgebras}~\cite{CanningFbounded89,Jacobs95,Poll2000276}.
As such, building on these earlier successes in the use of category theory 
in programming languages research, we intend to consider using \emph{operads}~\cite{spivak2014category}
to model the self-similarity (see page~\pageref{lbl:selfsim} above) of 
the subtyping relation in generic nominally-typed OOP.

We believe the inclusion of nominal information, the use of notions
such as nominal intervals and full generification, and the possible
use of category theory, may hold the key towards constructing a mathematical
model of generic nominally-typed OOP that, unlike prior mathematical
models of OOP, is simple yet accurate enough to provide a solid understanding
of the main features of generic nominally-typed OOP.

\bibliographystyle{plain}

\appendix

\section{\label{sec:Equations-for-Generic}Formal Definition of Generic Class Signatures}

Formally, akin to $\NOOP$ class signatures, $\GNOOP$ class signature constructors
are defined as
\begin{eqnarray}
\mathsf{SC} & = & \mathsf{N}\times\mathsf{X}^{*}\times\mathsf{GN}^{*}\times(\mathsf{L}\times\mathsf{GNX})^{*}\times(\mathsf{L}\times\mathsf{GNX}^{*}\times\mathsf{GNX})^{*}\label{eq:sig-cons}
\end{eqnarray}
where $\mathsf{SC}$ is the set of class signature constructors, $\mathsf{N}$
is the set of class/interface/trait names, $\mathsf{X}$ is the 
set of type variable names, $\mathsf{L}$ is the set of member (\emph{i.e.},
field/method) names, and $\times$ and $^{*}$ are the cross-product
and finite sequences set constructors.\footnote{As such, $\mathsf{X}^{*}$ 
is the set of (finite) sequences of type variables. As a component inside 
a class signature constructor, a
member of $\mathsf{X}^{*}$ is the sequence of type variables whose
members can be used inside this class signature constructor. Ordering
of elements in a sequence of type variables (\emph{i.e}.\emph{, }an
element of $\mathsf{X}^{*}$) does matter (type arguments are matched
with type variables based on the \emph{order} of each in their respective
sequences). Repetition is not allowed allowed in elements of $\mathsf{X}^{*}$.}

For `generic signature names' $\mathsf{GN}$, we have 
\[
\mathsf{GN}=\mathsf{N}\times\mathsf{GNX}{}^{*},
\]
and for `generic signature names or type variables' $\mathsf{GNX}$,
we have 
\[
\mathsf{GNX}=\mathsf{GN}+\mathsf{X}.
\]
(Note the \emph{mutual dependency} between $\mathsf{GN}$ and $\mathsf{GNX}$,
and that only members of $\mathsf{N}$, not $\mathsf{X}$, can be
paired with members of $\mathsf{GNX}{}^{*}$.)

\section{\label{sec:Fully-Generified-Generic-Class}Fully-Generified Generic
Class Signatures}

Formally, in an approach based on nominal intervals and full generification,
the equations for generic signatures will be as follows (in contrast
to using $\mathsf{X}$ in the equations above, $\mathsf{Y}$ in the
following equations is the set of \emph{both }synthetic and original
type variable names), where we have

\[
\mathsf{GN}=\mathsf{N}\times\mathsf{Y}{}^{*}
\]
\[
\mathsf{GNY}=\mathsf{GN}+\mathsf{Y}
\]
(Note the \emph{single-nesting} of generic types. There is \emph{no} mutual
dependency between $\mathsf{GN}$ and $\mathsf{GNY}$, as that between
$\mathsf{GN}$ and $\mathsf{GNX}$ above.)

For nominal intervals (modeling bounded type variables, and wildcard
types captured in synthetic type variables), we have

\[
\mathsf{YB}=\mathsf{Y}\times\mathsf{GNY}\times\mathsf{GNY}.
\]
(All type variables have lower and upper bounds, in addition to a
name. For a triple in $\mathsf{YB}$ to be a nominal interval the lowerbound
(the second component of the triple) must be a subtype of the upperbound
(the third component of the triple). Since they may stand for different
(unknown) types, nominal intervals with the \emph{same} lower and upper
bounds but with \emph{different} names are \emph{different }intervals---hence
nominality. Also, a non-circularity convention---as in Scala, and more recently
in Java\footnote{Based on the announced Java 7.0 language enhancements (at
http://docs.oracle.com/javase/7/docs/technotes/guides/language/enhancements.html\#javase7),
making Java follow the footsteps of other nominally-typed OO languages
(such as Scala) in allowing naked type variables occur as bounds of
earlier-declared type variables (\emph{i.e.}, having naked forward
references) seems to be a (surprising) \emph{unannounced} language enhancement
in Java 7.0 (released in 2011), even though the relaxation of this restriction
is mentioned in~\cite[Sec. 6.3]{JLS14} where it is mentioned that `the scope
of a class's type parameter is the type parameter section of the class
declaration ... '.

When generics were initially introduced
in Java 5.0 (in 2004), it was required that a type variable bound
\emph{not} be a naked type variable that is declared later in the
type parameters clause of a generic class/interface (possibly due
to an indirect influence from~\cite{Baldan1999}, which has the same
restriction). The original restriction was reflected, for example, in 
question TypeParameters.FAQ302 of (earlier versions of) the Java Generics FAQ~\cite{GenericsFAQWebsite}, where it is mentioned that the code
\begin{lyxcode}
	<S extends T, T extends Comparable<S>\.> T create(S arg) \{\ ...\ \}
\end{lyxcode}
which now successfully compiles in Java, is in error, and that the forward
reference (to type variable \code{T}) in the code is illegal.
 
The non-circularity convention of Scala---\emph{i.e.},
that a type parameter may not be bounded directly or indirectly by
itself---is explicitly mentioned in the Scala language specification,
at least as of Scala 2.7~\cite[Sec. 4.4]{Odersky09} (in 2009). Java
now seemingly (silently) follows suit and makes the scope of a type
parameter include the whole type parameter clause. Therefore, in Scala
and in the latest versions of Java, it is possible for a type parameter
to appear as part of its own bounds or the bounds of any of the other
type parameters in the same clause. (We thank Martin Odersky
for the brief discussion on this point.)
}---will enforce that a type variable
does \emph{not} occur by itself directly or indirectly in its own bounds.)

Then, for class signature constructors, we have

\begin{eqnarray*}
\mathsf{SC} & = & \mathsf{N}\times\mathsf{YB}^{*}\times\mathsf{GN}^{*}\times(\mathsf{L}\times\mathsf{Y})^{*}\times(\mathsf{L}\times\mathsf{Y}^{*}\times\mathsf{Y})^{*}
\end{eqnarray*}
(Note: In comparison with Equation~\eqref{eq:sig-cons}, \emph{all} types of fields and methods are now references
to (original or synthetic) type variables/nominal intervals. A class signature constructor
is thus now \emph{fully generified}).

\section{\label{sec:Full-Generification-Code-Examples}Full-Generification
Code Examples}

For an illustration of full generification, consider the simple generic 
class declaration
\begin{lyxcode}
\begin{lstlisting}[language=Java,tabsize=4,frame=top left right bottom,frameround=top]
class C<T> {
  private Integer count;
  private T t;
  
  C(T t) { this.t = t; count = 0; }
}
\end{lstlisting}

\end{lyxcode}
Using the notation \code{{[}VN:LB-UB{]}} for a nominal interval that
denotes a bounded type variable with name \code{VN},
a lower bound \code{LB} and an upper bound \code{UB} (and where
\code{O} and \code{N} are shorthands for a type \code{Object},
at the top of the subtyping heirarchy, and a type \code{Null} at
its bottom), the declaration of class \code{C}, when fully generified,
gets translated to
\begin{lyxcode}
\begin{lstlisting}[language=Java,tabsize=4,frame=top left right bottom,frameround=left]
// T is original, T1 to T3 are synthetic
class C<[T:N-O],[T1:Integer-Integer],
        [T2:T-T],[T3:T-T]> {
  private T1 count;
  private T2 t;
  
  C(T3 t) { this.t = t; count = 0; }
}
\end{lstlisting}
\end{lyxcode}

\hrule
\vskip 10pt

Using the same notation, when fully generified, the (non-generic) class declaration
\begin{lyxcode}
\begin{lstlisting}[language=Java,tabsize=4,frame=top left right bottom,frameround=top]
class DecorCanvas extends Canvas {
  void drawShapes(List<? extends Shape> ss){
    for(s: ss){
      s.draw(this);
      drawDecor(s);
    }
  }
}
\end{lstlisting}
\end{lyxcode}

gets translated to the (generic) class declaration

\begin{lyxcode}
\begin{lstlisting}[language=Java,tabsize=4,frame=top left right bottom,frameround=left]
// T1 and T2 are synthetic
class DecorCanvas<[T1:N-Shape],
  [T2:List<T1>-List<T1>]>
  extends Canvas {
  
  void drawShapes(T2 ss){
    for(s: ss){
      s.draw(this);
      drawDecor(s);
    }
  }
}
\end{lstlisting}
\end{lyxcode}

and the class declaration

\begin{lyxcode}
\begin{lstlisting}[language=Java,tabsize=4,frame=top left right bottom,frameround=top]
class ListCopier<T> {
  void copy(List<? extends T> src,
            List<? super T> dest){
    for(int i=0; i < src.size(); i++)
      dest.set(i, src.get(i));
  }
}
\end{lstlisting}
\end{lyxcode}

gets translated to

\begin{lyxcode}
\begin{lstlisting}[language=Java,tabsize=4,frame=top left right bottom,frameround=left]
// T is original, T1 to T4 are synthetic
class ListCopier<[T:N-O],
  [T1:N-T],[T2:List<T1>-List<T1>],
  [T3:T-O],[T4:List<T3>-List<T3>]> {
  void copy(T2 src, T4 dest){
    for(int i=0; i < src.size(); i++)
      dest.set(i, src.get(i));
  }
}
\end{lstlisting}
\end{lyxcode}

\hrule
\vskip 10pt

For a more intricate example of full generification, consider the declaration

\begin{lyxcode}
\begin{lstlisting}[language=Java,tabsize=4,frame=top left right bottom,frameround=top]
class Box<T> {
  private T t;
  Box(T t) { this.t = t; }
  void put(T t) { this.t = t;}
  T take() { return t; }
  boolean equalTo(Box<?> other) { 
    return this.t.equals(other.t); }
  Box<T> copy() { return new Box<T>(t); }
}
\end{lstlisting}
\end{lyxcode}

which, when fully generified, gets translated to

\begin{lyxcode}
\begin{lstlisting}[language=Java,tabsize=4,frame=top left right bottom,frameround=left]
// T is original, T1 to T11 are synthetic
class Box<[T:N-O],[T1:T-T],[T2:T-T],[T3:T-T],
  [T4:T-T],[T5:Boolean-Boolean],
  [T6:T-T], /* Not [T6:N-O], due to
             * "wildcard capturing" */
  [T7:Box<T6>-Box<T6>],[T8:T-T],
  [T9:Box<T8>-Box<T8>],
  [T10:T-T],[T11:Box<T10>-Box<T10>]> {
  private T1 t;
  Box(T2 t) { this.t = t; }
  void put(T3 t) { this.t = t;}
  T4 take() { return t; }
  T5 equalTo(T7 other) { 
    return this.t.equals(other.t); }
  T9 copy() { return new T11(t); } 
}
\end{lstlisting}
\end{lyxcode}

\end{document}